\documentclass[final,3p,times,twocolumn]{elsarticle}

\usepackage{amssymb}

\journal{Physica D}

\newcommand{\arcsec}{\mathrm{arcsec}}
\newcommand{\Amin}{A_\mathrm{min}}
\newcommand{\Amax}{A_\mathrm{max}}
\newcommand{\dimtemp}{\Delta}
\newcommand{\gbar}{h}
\newcommand{\Wbar}{W_\gbar}
\newcommand{\Vbar}{V_\gbar}

\begin{document}

\begin{frontmatter}



\title{Phase field modelling of grain boundary premelting using obstacle potentials}


\author[mpie]{V. Sai Pavan Kumar Bhogireddy}
\author[mpie]{C. H\"uter}
\author[mpie]{J. Neugebauer}
\author[icams]{I. Steinbach}
\author[neu]{A. Karma}
\author[mpie]{R. Spatschek}

\address[mpie]{Max-Planck-Institut f\"ur Eisenforschung GmbH, 40237 D\"usseldorf, Germany}
\address[icams]{Interdisciplinary Centre for Advanced Materials Simulation (ICAMS), Ruhr-Universit\"at Bochum, 44780 Bochum, Germany}
\address[neu]{Physics Department and Center for Interdisciplinary Research on Complex Systems, Northeastern University, Boston, MA 02115, USA}

\begin{abstract}
We investigate the multi-order parameter phase field model of Steinbach and Pezzolla [I. Steinbach, F. Pezzolla, A generalized field method for multiphase transformations using interface fields, Physica D 134 (1999) 385-393] concerning its ability to describe grain boundary premelting.
For a single order parameter situation solid-melt interfaces are always attractive, which allows to have (unstable) equilibrium solid-melt-solid coexistence above the bulk melting point.
The temperature dependent melt layer thickness and the disjoining potential, which describe the interface interaction, are affected by the choice of the thermal coupling function and the measure to define the amount of the liquid phase.
Due to the strictly finite interface thickness also the interaction range is finite.
For a multi-order parameter model we find either purely attractive or purely repulsive finite-ranged interactions.
The premelting transition is then directly linked to the ratio of the grain boundary and solid-melt interfacial energy.
\end{abstract}


\end{frontmatter}


\section{Introduction}
\label{intro::section}

Multi-phase field models are a powerful method to simulate complex microstructure evolution and interfacial pattern formation processes in a wide range of applications \cite{KarmaReview, ChenReview, steinbach, Steinbach:2013aa, Spatschek11}.
Whereas the role of isolated interfaces both in equilibrium and non-equilibrium is well understood, there is still a lack of understanding of the interaction of interfaces.
In a phase-field context, interactions appear when the smooth order parameter profiles with a width $\xi$ overlap.
These interactions can strongly influence the behaviour of polycrystalline materials, especially at elevated temperatures.
This interaction is relevant to understand phenomena like grain coalescence \cite{Rappaz:2003aa} and grain boundary premelting \cite{Mishin:2009aa}.
The latter has widely been studied experimentally \cite{Glicksman:1972aa, Hsieh19891637, Alsayed:2005aa} and in various modelling methods, among them lattice models \cite{Besold:1994aa,PhysRevB.21.1893}, molecular dynamics or Monte Carlo simulations \cite{PhysRevE.79.020601, Williams20093786}, phase field models \cite{PhysRevE.81.051601}, orientational order parameter phase field models \cite{Tang:2006aa,Lobkovsky2002202}, phase field crystal \cite{PhysRevB.78.184110,Adland:2013ys,PhysRevB.77.224114} and amplitude equations descriptions \cite{Spatschek:2010fk,Huter:2014aa,kar13}, with a strong influence on nucleation \cite{Frolov:2011fk} and melting kinetics \cite{Huter:2014uq}.
This progress established the understanding that in general high energy grain boundaries tend to premelt, in contrast to those with a small misorientation angle.
This effect is attributed to a short-ranged interaction between adjacent solid-melt interfaces, denoted as disjoining potential.
For an attractive potential, the grain boundary can even sustain temperatures above the melting point (provided that surface melting is inhibited), whereas repulsive interactions lead to the formation of a thin melt layer already below the bulk melting temperature $T_M$.
From an energetic standpoint one expects  that the ratio of the grain boundary energy $\sigma_{gb}$ to twice the solid-melt interfacial energy $2\sigma_{sl}$ is the decisive parameter at the melting point:
For $\sigma_{gb}/2\sigma_{sl}$ larger (smaller) than unity the dry grain boundary is energetically less (more) favourable, and we therefore expect a repulsive (attractive) interaction between the solid-melt interfaces.
Phenomenologically, the interaction is then described by an exponential decay \cite{Widom:1978aa,Rappaz:2003aa}.
We note that in general one has to carefully extrapolate the temperature-dependent grain boundary energy to the melting point to predict the premelting transition correctly \cite{Olmsted:2011vn}.
Another application of the above mentioned problem, besides grain boundary premelting, is the existence of thin $\gamma$ channels in between $\gamma'$ precipitates in Ni-base superalloys. 
In these alloys wetting of the $\gamma$ matrix between precipitates prevents topological inversion of matrix and precipitate phase. 
This will be treated in detail elsewhere.

The multi-order parameter phase field model \cite{Steinbach:1998aa} is frequently used for many materials science related questions and the basis for the phase field codes {\sc Micress} and {\sc OpenPhase} \cite{openphase}.
One of the benefits of this phase field model is that in a dual interface no undesired third phase contributions appear.
This is an advantage in comparison to simple multi-order parameter phase field models based on a multi-well potential, unless careful efforts are made to suppress spurious third-phase contributions in such dual interfaces \cite{Folch:2005kx}.
Here, in contrast, we are particularly interested in situations, where third phases (e.g.~a liquid) appear at the interface between two others (which can also be grains of the same material with a different orientation).
Already from a practical point of view it is important to understand the behaviour of this model if grain boundaries are brought to temperatures close to the melting point --- a situation which is naturally encountered in modelling of solidification.
The guiding question is therefore whether the model is appropriate to capture the grain boundary premelting effect in a phenomenological sense.
A qualitative inspection of this model has been discussed briefly for the single order parameter case in \cite{PhysRevE.81.051601}, but neither quantitative results were derived, nor the important extension to the multi-order parameter case has been pursued, which allows to have not only attractive interface interactions.
It is therefore the purpose of the present paper to shed light on the grain boundary premelting behavior of this multi-order parameter phase field model.

The article is organised as follows:
In section \ref{model::section} the phase field model is briefly presented.
Section \ref{single::section} analyses the model for a single order parameter case, where only one grain orientation is considered.
Also, there shall be no translational misfit between the grains, as discussed in \cite{kar13}.
In this case adjacent solid-melt interfaces always attract each other, in agreement with the expectation that the disappearance of the thin melt layer reduces the total interfacial energy.
The interaction range is strictly finite.
Moreover, we confirm that despite the mathematical nature of the model, which implies a piecewise solution of the phase field profiles, the order parameters do have a continuous slope at the transition points where the number of locally present phases changes.
In section \ref{multi:section} the results will be generalised to the multi-order parameter case, where the distinction between different grain orientations can be encoded by the phase fields.
Here, the interactions turn out to be either attractive or repulsive, depending on the ratio  of the dry grain boundary energy to the solid-melt interfacial energy.
Again, the order parameters are spatially continuous and kink-free.

\section{Model description}
\label{model::section}

Starting point of the description is a free energy functional \cite{Steinbach:1998aa},
\begin{eqnarray}
 F_0 &=& \int \Bigg\{ \sum_{\alpha=1}^N \sum_{\beta >\alpha}^N \Big(\frac{-4\eta_{\alpha\beta}\sigma_{\alpha\beta}}{\pi^2}\nabla\phi_\alpha \nabla\phi_\beta \nonumber \\
&+& \frac{4\sigma_{\alpha\beta}}{\eta_{\alpha\beta}}\phi_\alpha \phi_\beta \Big) \nonumber \\
&-&  \frac{L(T-T_M)}{T_M}[1-g(\{\phi_\alpha\})] \Bigg\} dV,  \label{intro::eq1}                                           
\end{eqnarray}
which depends on the (dimensionless) phase fields (or order parameters) $\phi_\alpha$.
Each of them characterises one phase, and in the bulk for the locally present phase $\phi_\alpha=1$, whereas all others vanish, $\phi_\beta=0$ for $\beta\neq \alpha$.
Here, $N$ is the maximum number of phases which may appear in the description.
With the constraint
\begin{equation}
\sum_\alpha^N \phi_\alpha=1 \label{intro::eq2}
\end{equation}
at each position, the phase fields may also be interpreted as local volume fractions of the phases.
In this sense it is imposed that the phase fields should be non-negative, $\phi_\alpha\geq 0$, hence by Eq.~(\ref{intro::eq2}) also $\phi_\alpha\leq 1$.
In the model formulation this is formally enforced by an infinite energy penalty if the phase field values leave the domain of allowed values, i.e.~the generating functional is
\begin{equation} \label{intro::eq3}
F = F_0 + \left\{ 
\begin{array}{cc}
0 & 0\leq \phi_\alpha\leq 1, \alpha=1, \ldots, N \\
\infty & \mbox{else}
\end{array}
\right. .
\end{equation}
The second term in the expression (\ref{intro::eq1}), i.e.~the local term $\phi_\alpha\phi_\beta$ is called the {\em obstacle potential}.
Together with the penalty terms in Eq.~(\ref{intro::eq3}) it leads to the fact that the phase fields become strictly one or zero outside an interface region.
In this sense the model differs from phase field formulations with a multi-well potential, where the phase fields typically have the form $\phi_\alpha(x)=(1+\tanh x)/2$, and therefore approach the bulk values exponentially but never strictly reach them.
For practical purposes the multi-obstacle potential therefore allows to rigorously confine the phase field evolution to the interface regions, which can be used to accelerate the simulations.
Further parameters appearing in Eq.~(\ref{intro::eq1}) are the interfacial energies $\sigma_{\alpha\beta}=\sigma_{\beta\alpha}$ (dimension: $\mathrm{J}/\mathrm{m}^2$) and the interface thickness parameters $\eta_{\alpha\beta}=\eta_{\beta\alpha}$ (dimension: $\mathrm{m}$).
The tilt function $g(\{\phi_i\})$, which may depend on several order parameters, shall be zero in the liquid and one in the solid, such that deviations of the (homogeneous) temperature $T$ from the melting temperature $T_M$ favour either the solid or liquid phase.
This term also contains the latent heat $L$ (dimension: $\mathrm{J}/\mathrm{m}^3$).

The dynamics of the phase fields is then expressed through interface fields
\begin{equation} \label{intro::eq4}
\dot{\psi}_{\alpha\beta} := - \left( \frac{\delta}{\delta \phi_\alpha} - \frac{\delta}{\delta \phi_\beta} \right) F,
\end{equation}
and for the phase field evolution in the interface regions, $0< \phi_\alpha < 1$
\begin{equation} \label{intro::eq5}
\dot{\phi}_\alpha = \frac{1}{\tilde{N}} \sum_{\beta\neq \alpha} \mu_{\alpha\beta} \dot{\psi}_{\alpha\beta},
\end{equation}
with kinetics coefficients $\mu_{\alpha\beta}=\mu_{\beta\alpha}>0$.
Here, $\tilde{N}$ is the number of phases with non-vanishing volume fractions at the present position.

The above structure of the model formulation indicates that the fields should be determined piece-wise, and therefore also the interface profiles will depend sensitively on the number of locally present phases.
In the context of grain boundary wetting we therefore carefully have to distinguish between single phase regions, dual interfaces and regions, where additionally to solid phases a melt appears.

In the present paper we focus exclusively on equilibrium properties, thus in the above equation (\ref{intro::eq5}) the time derivative on the left hand side vanishes.
We have shown in a previous publication that the stationarity of the phase fields also implies that interface fields vanish, $\dot{\psi}_{\alpha\beta}=0$, as defined in Eq.~(\ref{intro::eq4}) \cite{Guo:2011aa}.
The resulting equilibrium conditions for the phase field can then also be obtained via energy minimisation.
First, let us consider a situation where all phase fields are non-trivial (i.e.~$0<\phi_\alpha<1$ for all $\alpha$), such that $F=F_0$.
To impose the constraint (\ref{intro::eq2}) we write
\begin{equation}
\tilde{F}(\phi_1, \ldots, \phi_{N-1}) = F\left( \phi_1, \ldots, \phi_{N-1}, 1-\sum_{\alpha=1}^{N-1}\phi_\alpha \right).
\end{equation}
Hence the energy minimisation conditions read
\begin{equation}
\frac{\delta \tilde{F}}{\delta \phi_\alpha} = \frac{\delta F}{\delta \phi_\alpha} - \frac{\delta F}{\delta \phi_N} = -\dot{\psi}_{\alpha N}=0
\end{equation}
for $\alpha=1,\ldots, N-1$.
Noting further that $\dot{\psi}_{\alpha\beta} = \dot{\psi}_{\alpha N} - \dot{\psi}_{\beta N}$, we can therefore conclude that a vanishing right hand side of Eq.~(\ref{intro::eq4}) for all pairs $\alpha, \beta$ is equivalent to the minimisation of the free energy under the constraint (\ref{intro::eq2}).
Here we point out, that these conditions only hold provided that all phase fields are non-trivial, as then the equilibrium is determined by a local extremum of the functional $F$.
If, however, at least one phase field becomes zero or one, the discontinuity of the formal energy penalty term in Eq.~(\ref{intro::eq3}) leads to a solution via a global minimum of the functional, and consequently the local conditions $\dot{\psi}_{\alpha\beta}=0$ are inconvenient to apply.
Here, instead, the energy minimisation picture is more useful.
To illustrate this we use the specific case $N=3$ which is most relevant for the present paper.
Let us assume that phase 3 is not appearing, i.e.~$\phi_3\equiv 0$.
Then we write the energy functional as $\bar{F}(\phi_1)=F(\phi_1, 1-\phi_1, 0)$, and the equilibrium condition reads 
\begin{equation} \label{intro::eq8}
\frac{\delta \bar{F}}{\delta \phi}=0,
\end{equation}
with $\phi=\phi_1$ for brevity.
Hence the description effectively reduces to a single order parameter model, and the functional reads
\begin{eqnarray}
\bar{F}(\phi) &=& \int \Bigg\{ \frac{4\eta\sigma}{\pi^2}(\nabla\phi)^2 + \frac{4\sigma}{\eta}\phi(1-\phi) \nonumber \\
&& - L\frac{T-T_M}{T_M} [1-\gbar(\phi)]\Bigg\} dV, \label{intro::eq9}
\end{eqnarray}
in the interface region, where we defined $\sigma=\sigma_{12}$, $\eta=\eta_{12}$.
Furthermore, $\gbar(\phi)=g(\phi, 1-\phi, 0)$.

\section{Single order parameter description}
\label{single::section}

In this section we will focus on the single order parameter case, where we denote by $\phi=1$ the solid and by $\phi=0$ the liquid phase.
We investigate situations, where the planar liquid phase is sandwiched between two solid semi-infinite crystals.
This way, the description becomes one-dimensional.
If the melt layer thickness is large in comparison to the phase field interface thickness, stationary solutions only exist at $T=T_M$, due to the absence of an interface curvature.
If the interfaces come closer to each other, and they start to interact, hence a stationary solution can exist only with a different temperature.
In the single order parameter case, the two ``grains'' are physically identical, and therefore a complete merging, when the melt is removed, therefore leads to a perfect single crystal.
Apparently, this is an energetically favourable situation, and therefore we expect for the single order parameter case only attractive interactions between the solid-melt interfaces.
Nevertheless, the detailed investigation of this case is an important prerequisite for the multi-order parameter case in the following section, where also repulsive interactions can emerge.

\begin{figure}
\begin{center}
\includegraphics[width=8cm]{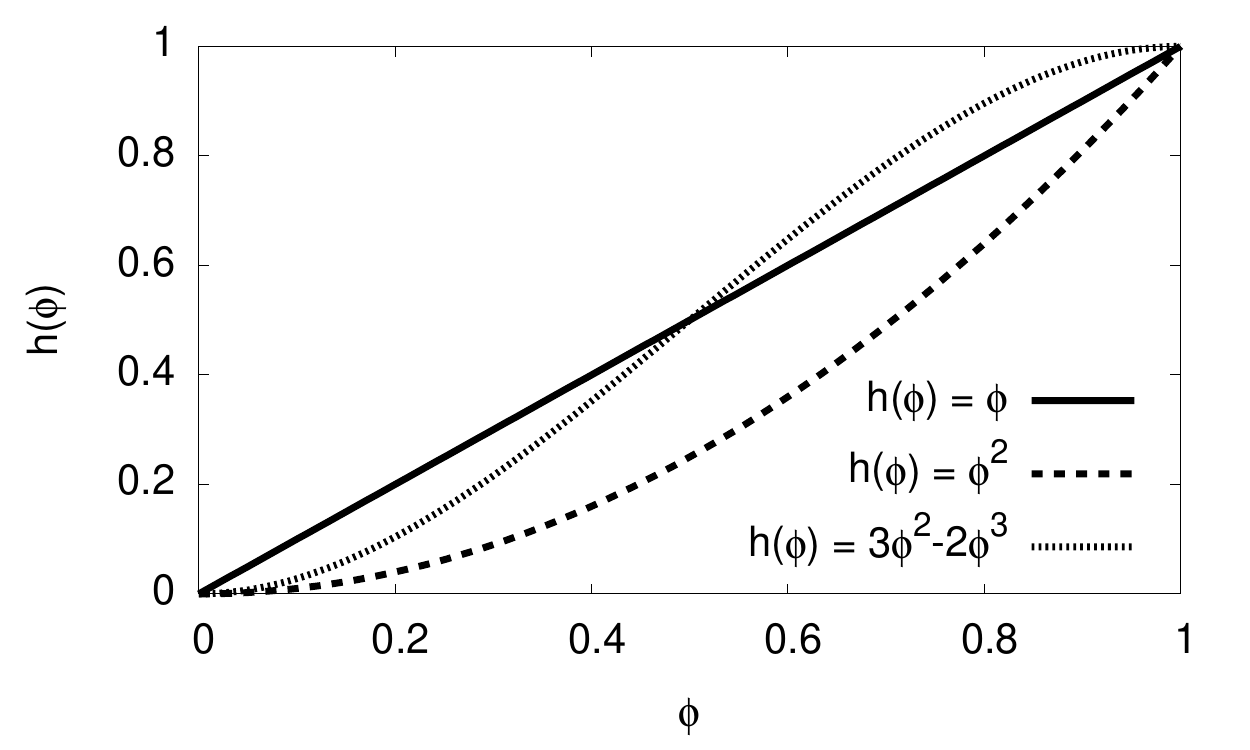}
\caption{Different choices of the coupling function $h(\phi)$ used in the paper.}
\label{fig7}
\end{center}
\end{figure}

In the interface region ($0<\phi<1$), the equilibrium condition reads according to Eqs.~(\ref{intro::eq8}) and (\ref{intro::eq9})
\begin{equation} \label{single::eq1}
-2\xi^2 \phi''(x) + 1-2\phi + \dimtemp\cdot \gbar'(\phi)=0,
\end{equation}
with the interface thickness $\xi=\eta/\pi$ and the dimensionless deviation from the melting temperature $\dimtemp=L\eta(T-T_M)/(4\sigma T_M)$. 
The prime denotes the spatial derivative in interface normal direction $x$.
The solution of this equation connects to the domains where $\phi=1$ in the solid phases.
Explicit solutions depend on the choice of the coupling function, and different cases will be discussed below, see Fig.~\ref{fig7}.
In this representation the free energy per unit area becomes
\begin{eqnarray} 
{\cal F} &=& \frac{4\sigma}{\eta} \int \Big\{ \xi^2(\phi')^2 + \phi(1-\phi) \nonumber \\
&&- \dimtemp\cdot [1-\gbar(\phi)] \Big\}\,dx. \label{single::eq1a}
\end{eqnarray}

A specific property of interest will be the thickness $W$ of the melt layer as function of temperature.
This quantity can be defined in different ways, and we use here two different measures for $W$.
The first is to call the system to be liquid, if this is the locally dominant phase, i.e.~$\phi<1/2$.
Hence we would define the melt layer thickness as the distance between the points where the phase field crosses this limiting value $\phi=1/2$.
Another way to define the melt layer thickness is to use the integral expression
\begin{equation} \label{single::eq2}
\Wbar = \int_{-\infty}^\infty [1-\gbar(\phi)]\,dx,
\end{equation}
which expresses that the amount of liquid is proportional to the latent heat contribution.
Both expressions deliver different results, but the qualitative picture remains unaffected.

Multiplication of Eq.~(\ref{single::eq1}) with $\phi'(x)$ and integration yields
\begin{equation} \label{single::eq3}
\xi^2[\phi'(x)]^2 - \phi(1-\phi) - \dimtemp\cdot \gbar(\phi) = -\dimtemp,
\end{equation}
where the integration constant on the right hand side results from notion that in the pure solid $\phi\equiv1$ the coupling function is $\gbar(1)=1$.
This equation can be interpreted as the representation of the phase field problems in the sense of the mechanical motion of a point particle in a one dimensional energy landscape.
For this, we have to read the spatial coordinate $x$ as time variable and consequently the spatial derivative as time derivative. 
The phase field value $\phi(x)$ is the position of the particle.
Then, the first term in the expression (\ref{single::eq3}) is the kinetic energy of the system, the other two terms the potential energy, and the right hand side $-\dimtemp$ is the conserved total energy.
In this picture, which leads to the ``Newtonian equation of motion'' (\ref{single::eq1}), the particle starts its motion from the position $\phi=1$ with vanishing velocity, rolls down in the potential energy landscape and accelerates.
It then rolls up to the classical turning point $\phi^*$, where $\phi'(x)=0$, i.e.~the particle shortly comes to rest and hence the kinetic energy vanishes.
This point is therefore defined by the condition
\begin{equation} \label{single::eq4}
\phi^*(1-\phi^*) = \dimtemp[1-\gbar(\phi^*)],
\end{equation}
which is therefore temperature dependent.
Since $0\leq \gbar\leq 1$ this equations has valid solutions $0<\phi^*<1$ only for $\dimtemp>0$, hence for overheating of the grain boundary.
This is reflecting the aforementioned attractive interaction of the solid-melt interfaces.
To compensate the attractive interaction between the solid-melt interfaces, the enclosed melt has to be stabilised by the overheating above the melting temperature.
We set the origin of the ``time'' $x=0$ exactly at the turning point $\phi=\phi^*$.
After that, the particle inverts its trajectory and moves back towards $\phi=1$.
Then by time inversion symmetry, $\phi(x)=\phi(-x)$.
For an extended discussion of this mechanical analogy we refer to \cite{PhysRevE.81.051601}.
In the mechanical interpretation the melt layer thickness is then nothing else than the ``time'' which the particle spends between the position $\phi=\phi^*$ and $\phi=1/2$.
Solving Eq.~(\ref{single::eq3}) for $\phi'(x)$ and integration therefore yields
\begin{equation} \label{single::eq5}
W(\dimtemp) = 2\xi \int_{\phi^*(\dimtemp)}^{1/2} \frac{d\phi}{\sqrt{\phi(1-\phi) + \dimtemp [\gbar(\phi)-1]} },
\end{equation} 
where the prefactor $2$ stems from the ``time'' inversion symmetry. 

From a thermodynamic perspective we can then represent the free energy of the system as 
\begin{equation} \label{single::eq6}
{\cal F}=-\frac{L(T-T_M)}{T_M} W + V(W) + 2\sigma,
\end{equation}
which first consists of the bulk term due to the deviation from the melting temperature;
this term is proportional to the melt layer thickness $W$.
Second, the disjoining potential $V(W)$ reflects the interface interaction.
Third, at large distances $W\ll\xi$, where the interaction has decayed, $V(W)\to 0$, only two solid-melt interfaces remain with interfacial energy $\sigma_{sl}=\sigma$ each.
In equilibrium $F'(W)=0$, hence from the inverted expression $\dimtemp=\dimtemp(W)$ we get for the disjoining potential
\begin{equation} \label{single::eq7}
V(W) = - \frac{4\sigma}{\eta} \int_W^\infty \dimtemp(W) dW.
\end{equation}

For the alternative definition of the melt layer thickness $\Wbar$ according to Eq.~(\ref{single::eq2}) we get
\begin{equation} \label{single::eq8}
\Wbar = \frac{1}{\dimtemp} \int_{-\infty}^\infty \left[ \phi(1-\phi) - \xi^2 (\phi')^2 \right]\, dx,
\end{equation}
which requires the knowledge of the phase field profile $\phi(x)$. 
The disjoining potential $\Vbar(\Wbar)$ is then defined as excess contribution
\begin{eqnarray}
\Vbar &=& {\cal F} + \frac{L(T-T_M)}{T_M} \Wbar - 2\sigma \nonumber \\
&=&  \frac{4\sigma}{\eta} \int\limits_{-\infty}^{\infty} \left\{ \xi^2(\phi')^2 + \phi(1-\phi) \right\}\,dx  - 2\sigma,
\end{eqnarray}
which vanishes for two non-interacting solid-melt interfaces.

To make the preceding steps more transparent, we use now three different coupling functions, see Fig.~\ref{fig7}, and derive explicit results for the melt layer thickness as function of temperature and the disjoining potential.
We mention for completeness, that there exists additionally one distinct coupling function which guarantees the so called ``traveling wave solution'' for a planar interface (see Appendix II in \cite{steinbach}). 
In the present work we restrict ourselves to power law functions in order to derive compact analytical expressions.
The generalisation to other coupling functions is straightforward.

\subsection{Case 1: $\gbar(\phi)=\phi$}
\label{case1::section}

The general solution of the stationary phase field equation (\ref{single::eq1}), which obeys the symmetry $\phi(x)=\phi(-x)$ is in the two-phase region
\begin{equation} \label{case1::eq1}
\phi(x) = \frac{\dimtemp+1}{2} - A \cos(x/\xi) + B \sin(x/\xi).
\end{equation}
with integration constants $A, B$.
Symmetry implies $B=0$.
For the moment we do not yet make assumptions on the value of $A$.
However, since outside the interface region $\phi=1$ we can conclude $A\geq \Amin$ with $\Amin=(1-\dimtemp)/2$.
In case of equality the phase field approaches the constant value smoothly, i.e. with continuous derivative $\phi'(x)$, whereas is all other cases it would exhibit a kink at the connection point.
Although we can immediately assume continuity of the phase field (otherwise the free energy would have a singular term through the gradient energy, and this would not be an equilibrium situation), continuity of the derivative is not directly obvious.
In fact, we have encountered in \cite{Guo:2011aa} situations with moving fronts where kinks naturally appear in the phase field profiles.
On the other hand, the phase field must not become negative, in agreement with the penally term (\ref{intro::eq3}), hence $A\leq \Amax$ with $\Amax=(1+\dimtemp)/2$.
Here one has to keep in mind that $\dimtemp>0$ due to the attractive interaction between the interfaces.

To shed light on the question whether kinks appear in the phase field profiles we calculate the free energy ${\cal F}$ as function of $A$.
The interface region is located between $|x|<x_0$ with $x_0=\xi \arccos[(\dimtemp-1)/2A]$, which is also the integration interval for the free energy expression (\ref{single::eq1a}).
Using the solution (\ref{case1::eq1}) then gives
\begin{equation}
{\cal F} = \frac{(1-\dimtemp)^2\xi }{2} \psi - A^2 \xi \sin(2\psi)
\end{equation}
with $\psi=\arcsec[2A/(\dimtemp-1)]$.
Equilibrium demands global minimisation of the expression ${\cal F}$ with respect to $A$.
This function has a local and global maximum at the upper bound $\Amax$ and a global minimum at $\Amin$.
Hence we can conclude that in equilibrium $A=\Amin$, which corresponds to the kink free solution.
Solution (\ref{case1::eq1}) is then valid in the regime $|x/\xi|\leq \pi$, and outside this interval $\phi=1$.
The absence of a kink is in agreement with the mechanical analogy, that the particle starts rolling down the negative obstacle potential coming from the rest position $\phi=1$.
Initially, the velocity of the particle is zero, hence $\phi'(-x_0)=0$, and this is already reflected in equation ({\ref{single::eq3}}) by proper choice of the integration constant.

\begin{figure}
\begin{center}
\includegraphics[width=8cm]{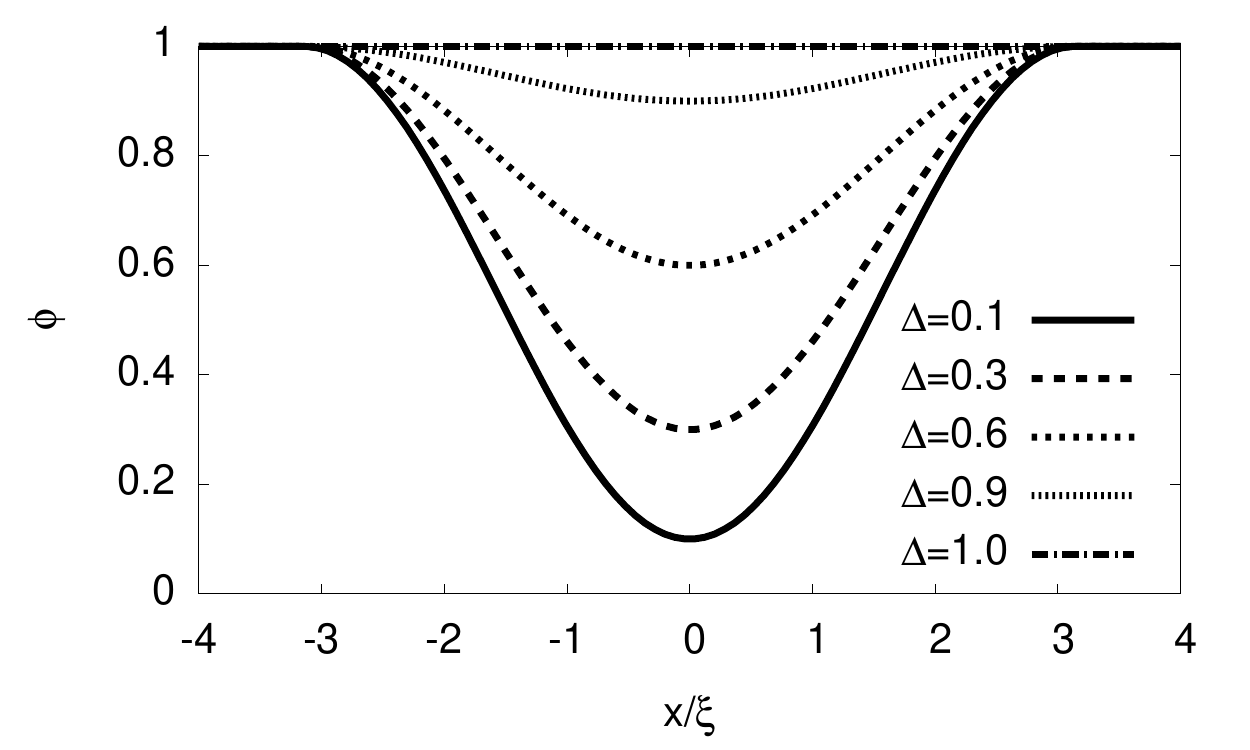}
\caption{Phase field profiles for different temperatures $\dimtemp$ in the single order parameter model with $\gbar(\phi)=\phi$.
Outside the interface region $|x|>\pi\xi$ the phase field is equal to one. At the connecting points $|x|=\pi\xi$ the profiles have a continuous slope.}
\label{fig1}
\end{center}
\end{figure}

Fig.~\ref{fig1} shows phase field profiles for different temperatures.
Closer to the melting point the melt layer thickness is higher.
We note that these profiles correspond to unstable solutions, i.e.~a maximum of the free energy (\ref{single::eq6}), hence small perturbations leads either to complete melting to disappearance of the melt layer (thus $\phi\equiv 1$ then).
For $\dimtemp>1$ the melt disappears completely, and then the phase field is constantly $\phi(x)\equiv 1$.

\begin{figure}
\begin{center}
\includegraphics[width=8cm]{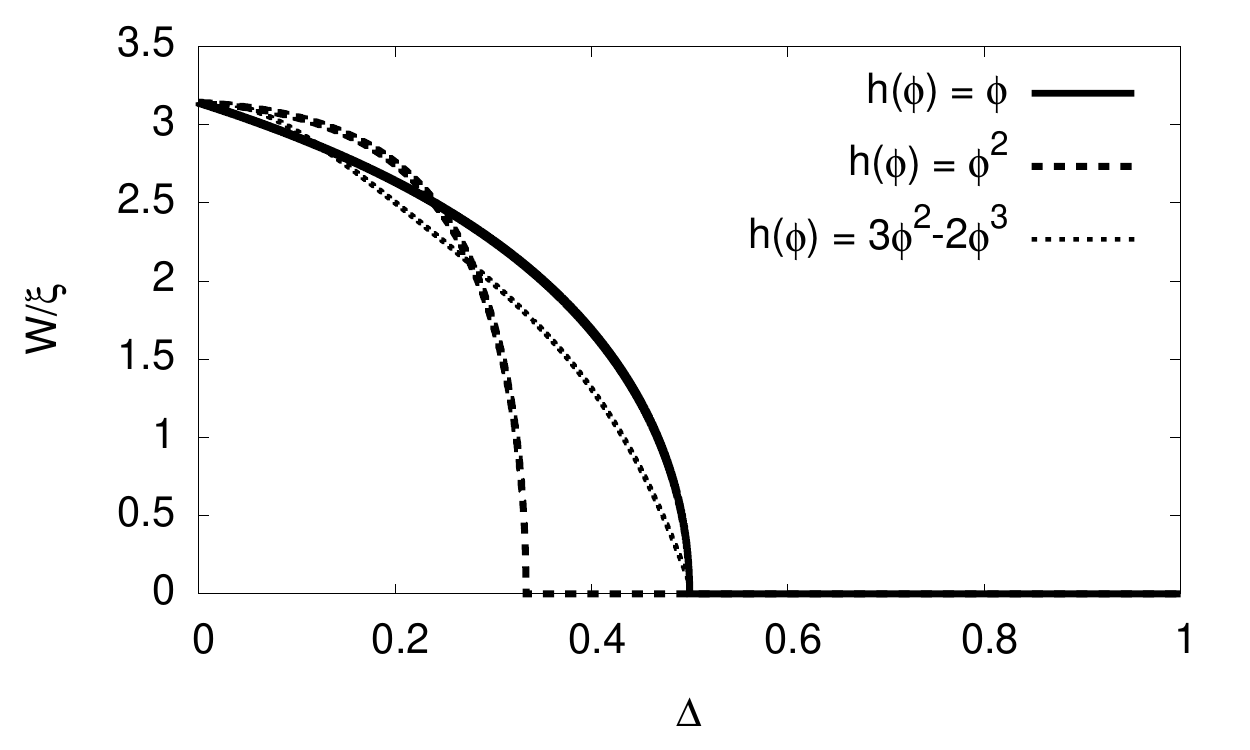}
\caption{Melt layer thickness as function of temperature for the single order parameter model with different choices of the coupling function.
}
\label{fig2}
\end{center}
\end{figure}

Using Eq.~(\ref{single::eq5}) we can then determine the equilibrium melt layer thickness as function of temperature.
Integration yields readily
\begin{equation}
W(\dimtemp) = 2\xi \left( \frac{\pi}{2} + \arcsin\frac{\dimtemp}{\dimtemp-1}\right),
\end{equation}
which  is shown in Fig.~\ref{fig2}.
For $\dimtemp>1/2$ the melt layer tickness $W$ becomes zero, which results from the definition of this measure.
As one can see in the interface profiles in Fig.~\ref{fig1} there is still a remaining volume fraction of the liquid, but since then $\phi>1/2$ it does not contribute to $W$.
On the other hand, if $\dimtemp$ approaches the melting point, the thickness still remains finite.
This is different from the phase field model with a double-well potential, where $W$ diverges logarithmically, if $T_M$ is approached from above \cite{PhysRevE.81.051601}.
Here, however, the two solid-melt interfaces do not overlap for $W>\pi\xi$, and therefore they do not interact.
Hence, such a solid-liquid-solid arrangement can only exist at $\dimtemp=0$.

\begin{figure}
\begin{center}
\includegraphics[width=8cm]{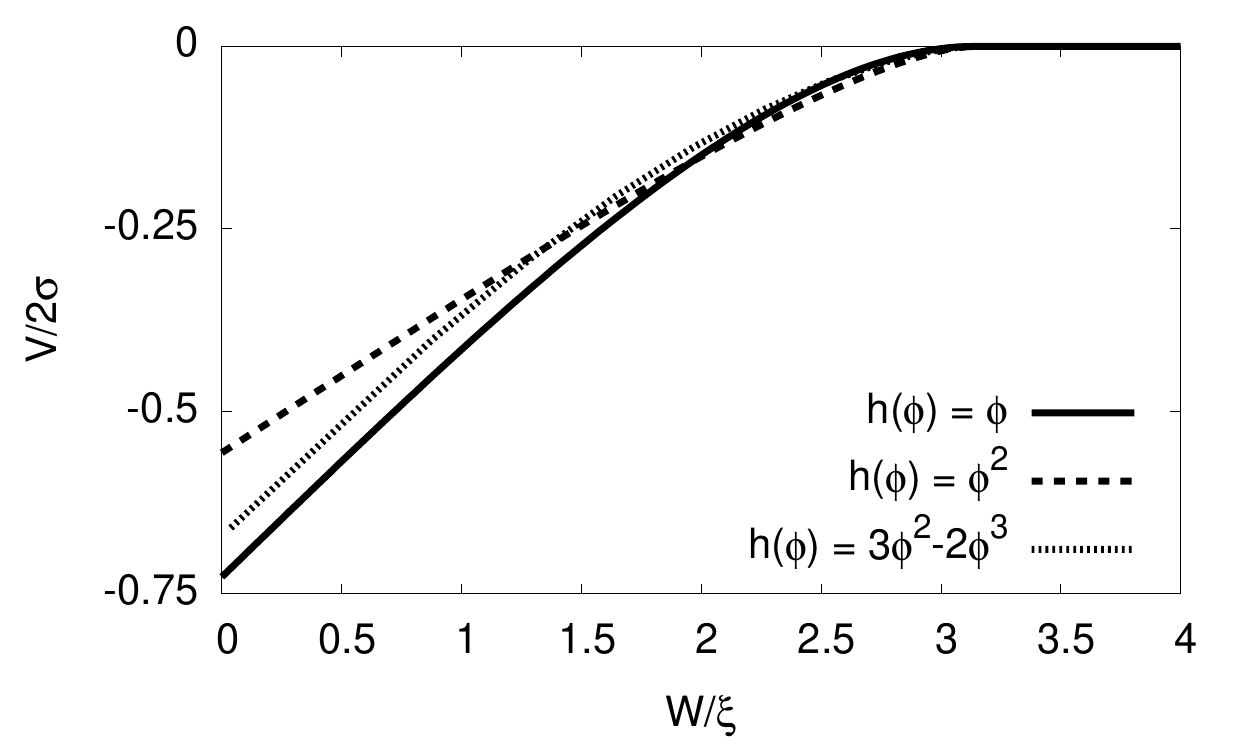}
\caption{Disjoining potential $V(W)$ for the single order parameter model for different choices of the coupling function.
In the case $\gbar(\phi)=\phi$ the disjoining potential is given by an analytical expression, in the other cases it is integrated numerically.
}
\label{fig3}
\end{center}
\end{figure}

Integration of the disjoining potential gives according to expression (\ref{single::eq7})
\begin{equation}
V(W) =  \frac{8\sigma}{\pi} \left[ \frac{W}{2\xi} - \tan\left({\frac{W}{4\xi}}\right) + 1- \frac{\pi}{2} \right],
\end{equation}
which is valid for $W\leq\pi\xi$ and purely attractive there, see Fig.~\ref{fig3}.
For $W$ larger than this value, the interaction vanishes.

\begin{figure}
\begin{center}
\includegraphics[width=8cm]{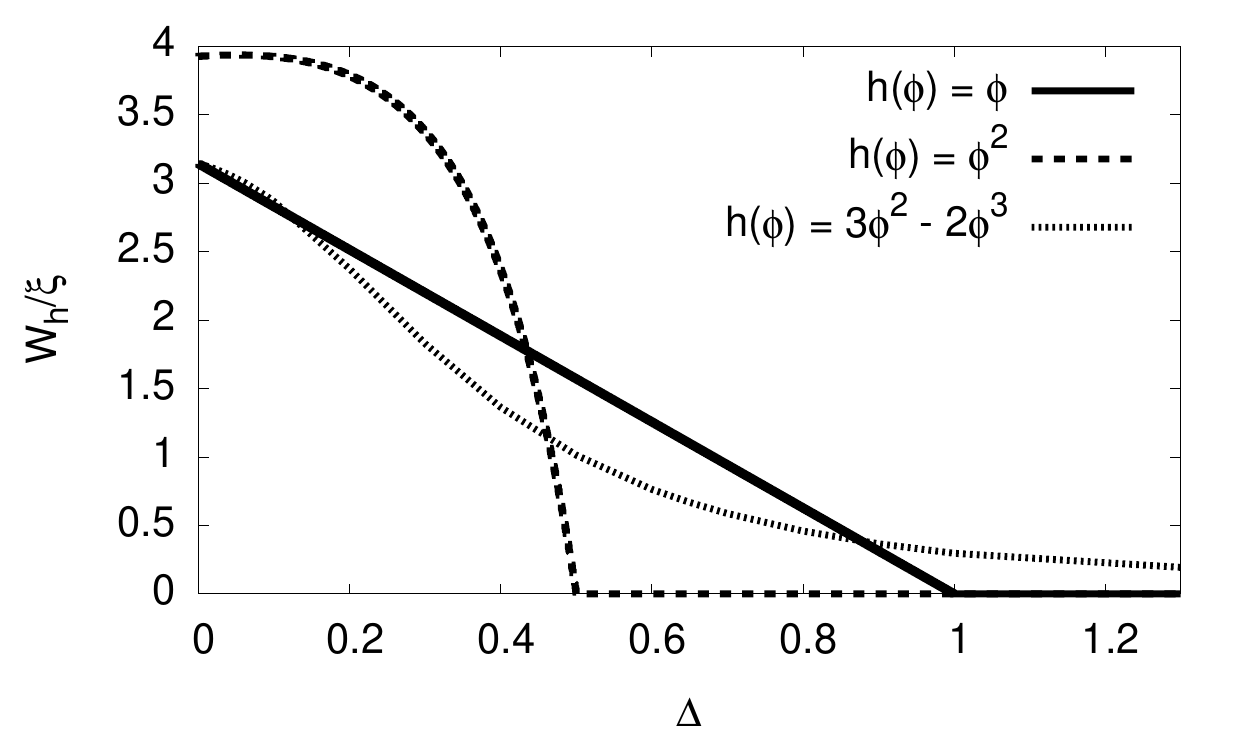}
\caption{Interface thickness for the alternative definition of the interface thickness $\Wbar$ for the different choices of the coupling function $\gbar(\phi)$ in the single order parameter model.
}
\label{fig4}
\end{center}
\end{figure}

If we use the alternative definition of the melt layer thickness (\ref{single::eq2}) we obtain from the solution (\ref{case1::eq1})
\begin{equation}
\Wbar(\dimtemp) = \pi \xi (1-\dimtemp),
\end{equation}
which is plotted in Fig.~\ref{fig4}.
In comparison to $W$ the liquid disappears here only for $\dimtemp\geq 1$.
Notice that beyond this threshold the phase field is therefore constantly $\phi(x)\equiv 1$.
This feature can easily be understood using the mechanical interpretation of the rolling particle in the negative potential energy landscape $U(\phi):=-\phi(1-\phi)-\Delta\cdot h(\phi)$, where the initial position of the particle is $\phi=1$.
The slope $U'(1)=1-\Delta\cdot h'(\phi)$ becomes negative for $\Delta>1$ for the present choice of the coupling function, and then the particle cannot leave the position $\phi=1$, as it would have to roll uphill.
Therefore, beyond this threshold no liquid phase can exist also in the spirit of $\Wbar$.

\begin{figure}
\begin{center}
\includegraphics[width=8cm]{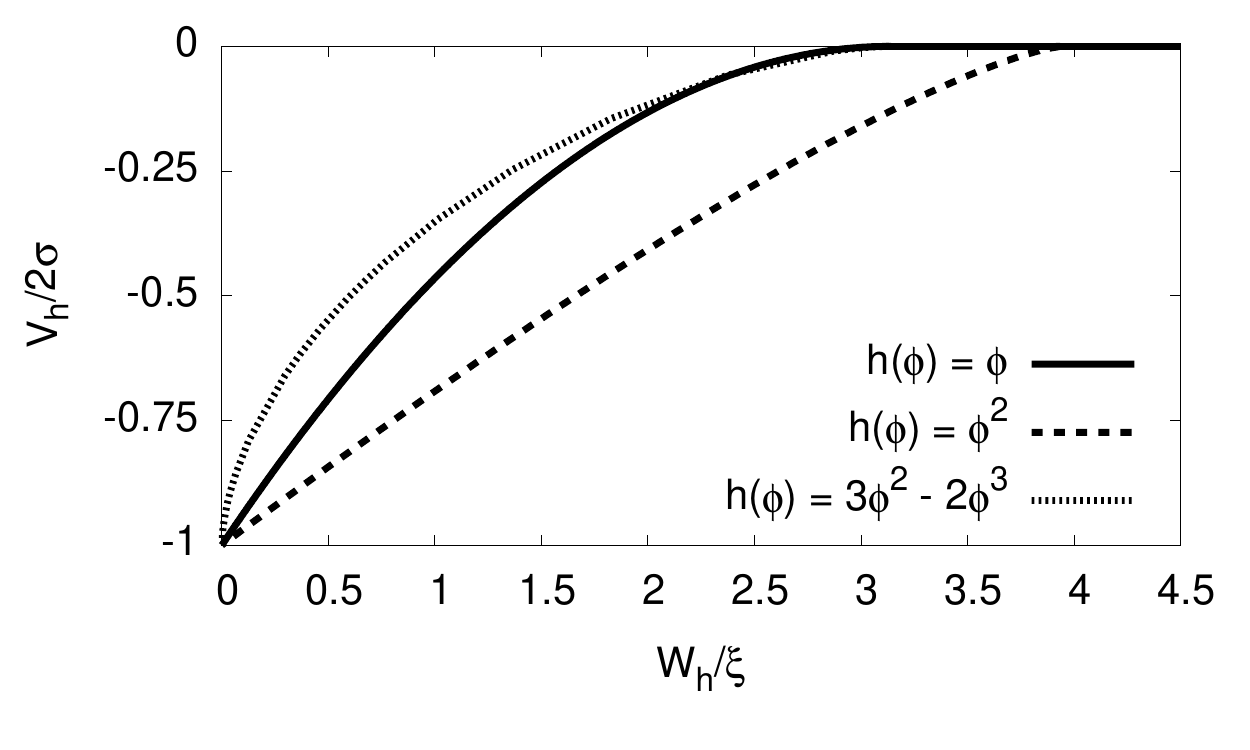}
\caption{Disjoining potential using the excess quantities $\Wbar$ and $\Vbar$ for the single order parameter model.
}
\label{fig5}
\end{center}
\end{figure}

The corresponding expression for the disjoining potential becomes
\begin{equation}
\Vbar(\Wbar) = -2\sigma \left( 1-\frac{\Wbar}{\pi\xi} \right)^2
\end{equation}
for $W\leq \pi\xi$ and $\Vbar=0$ otherwise.
This result is visualised in Fig.~\ref{fig5}.
With this definition, the interfacial energy becomes $2\sigma$ for fully separated solid-melt interfaces and vanishes for a perfectly healed crystal with $\Wbar=0$.
It therefore interpolates naturally between the two limiting values.

These results show that the results depend in detail on the precise definition of the observables, but the qualitative picture remains unchanged.
In particular, the interaction is attractive and vanishes for non-overlapping interfaces.

\subsection{Case 2: $\gbar(\phi)=\phi^2$}
\label{case2:section}

For this coupling function the equilibrium phase field profile reads in the interface region [i.e.~$|x/\xi|\leq \pi/\sqrt{1-\dimtemp}$]
\begin{equation}
\phi(x)=\frac{1}{2(1-\dimtemp)} - \frac{1-2\dimtemp}{2(1-\dimtemp)} \cos\frac{\sqrt{1-\dimtemp}x}{\xi}.
\end{equation}
Here, the integration constant is already adjusted such that the slope of the phase field is continuous at the border of the interface region.
Apparently, with this choice of $\gbar(\phi)$ the interface is stretched for deviations from bulk phase equilibrium, $\dimtemp=0$.
The same stretching factor appears in the melt layer thickness,
\begin{equation}
W(\dimtemp) = \frac{2\xi}{\sqrt{1-\dimtemp}} \left( \frac{\pi}{2} + \arcsin\frac{\dimtemp}{2\dimtemp-1} \right),
\end{equation}
which is shown in Fig.~\ref{fig2}.
At $\dimtemp=0$ the melt layer thickness is independent of the coupling function, and therefore the curves for different choices of $\gbar(\phi)$ come together at this point, where the two solid-melt interfaces just touch and do not overlap.
For higher temperatures, the curves $W(\dimtemp)$ differ for the different coupling functions. In particular, for the present choice $\gbar(\phi)=\phi^2$ the melt layer thickness becomes zero already at $\dimtemp=1/3$.

Here, the disjoining potential cannot be obtained in closed form and is therefore integrated numerically.
The result is shown in Fig.~\ref{fig3}.
As expected, it is close to the curve for the previous coupling function $\gbar(\phi)=\phi$, hence the choice of the coupling function is only of minor relevance.
The most important features are therefore that the interaction is -- as expected -- purely attractive, and that it has only a finite range.

For the alternative definition of $\Wbar$ and $\Vbar$ we get
\begin{equation}
\Wbar(\dimtemp) = \frac{\pi(2\dimtemp-5)(2\dimtemp-1)}{4(1-\dimtemp)^{5/2}}\xi,
\end{equation}
see Fig.~\ref{fig4}.
Due to the different definition the melt layer thickness of two non-overlapping interfaces is therefore $\Wbar(0)=5\pi\xi/4$.
In agreement with the mechanical picture the melt layer thickness is zero for $\Delta>1/2$, since then $U'(1)<0$.
Numerical inversion $\dimtemp(\Wbar)$ and insertion into the expression
\begin{equation}
\Vbar(\dimtemp) = \sigma \left( -2 + \frac{(1-2\dimtemp)(2 -\dimtemp + 2\dimtemp^2)}{(1-\dimtemp)^{5/2}} \right)
\end{equation}
delivers then the disjoining potential as function of the melt layer thickness, see Fig.~\ref{fig5}.


\subsection{Case 3: $\gbar(\phi)=3\phi^2-2\phi^3$}
\label{case3:section}

This choice of the coupling function has a certain popularity in models which use a double-well potential.
The reason is that apart from the properties $\gbar(0)=0$ and $\gbar(1)=1$ also $\gbar'(0)=\gbar'(1)=0$.
This has the consequence that the bulk states $\phi=0$ and $\phi=1$ are not shifted to different values for $\dimtemp\neq 0$ in these models.
For the obstacle potential this property is less crucial, as there the bulk state minimum in the potential landscape is cusp-like, and the thermal tilt does not shift the bulk states also for other choices of the coupling function.
Here we mainly use this function as an example for a case, where neither the phase field profile, the melt layer thickness nor the disjoining potential can be calculated analytically, and instead numerical integrations have to be used.
The results are shown in the preceding plots Figs.~\ref{fig2} and \ref{fig3}.
In particular, the disjoining potential is essentially in between the two previous cases.
Altogether we find that the disjoining potential is only rather weakly affected by the choice of the coupling function.



\begin{figure}
\begin{center}
\includegraphics[width=8cm]{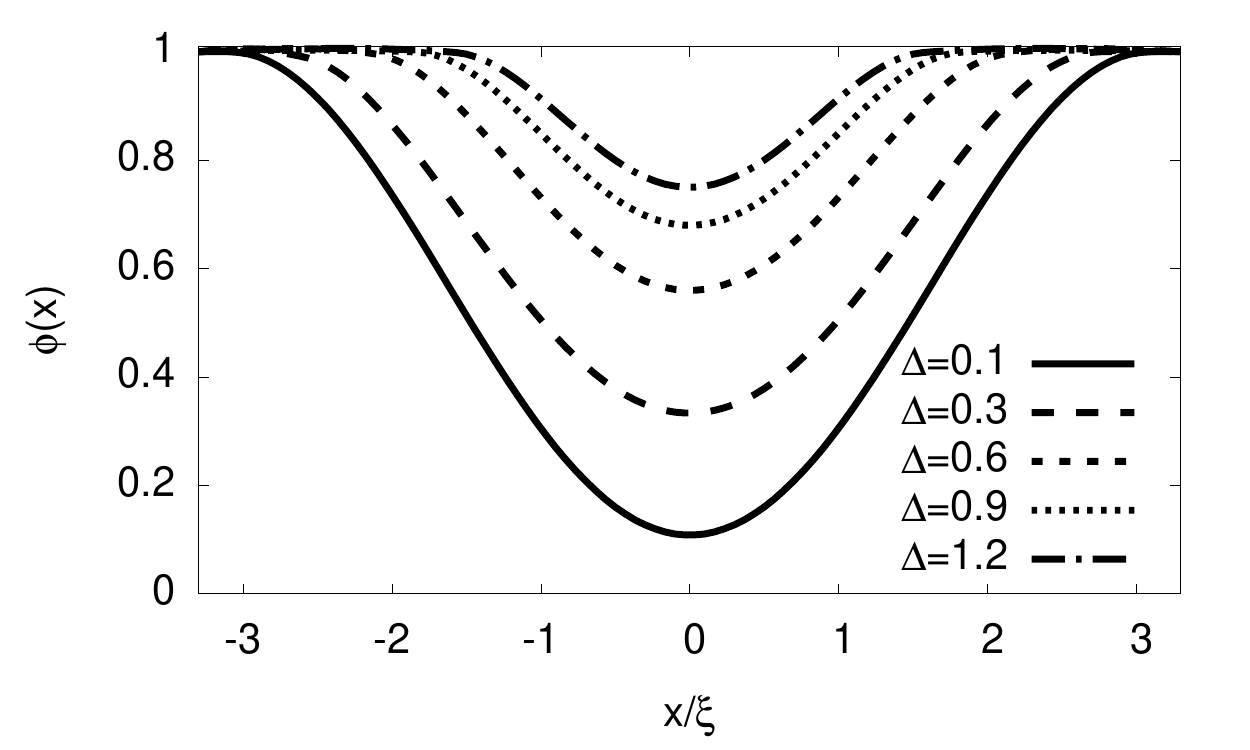}
\caption{Phase field profiles for different temperature for the coupling function $\gbar(\phi)=3\phi^2-2\phi^3$ in the single order parameter case.
}
\label{fig12}
\end{center}
\end{figure}

According to this definition of $\Wbar$ the solid-melt coexistence goes up to arbitrarily high temperatures, see Fig.~\ref{fig4}.
Due to the choice of the coupling function $\gbar'(1)=0$, and therefore $U'(1)>0$ for all $\Delta>0$, hence in the mechanical interpretation the particle always rolls a bit in the direction of smaller $\phi$ and therefore $\Wbar$ remains finite.
This is also visible in the phase field profiles in Fig.~\ref{fig12}.
In contrast to Fig.~\ref{fig1} here not only the drop of the phase field value decays with increasing temperature, but also the end points of the interface region move closer together.
Finally, the disjoining potential $\Vbar(\Wbar)$ is shown in Fig.~\ref{fig5}.

\section{Multi-order parameter description}
\label{multi:section}

\begin{figure}
\begin{center}
\includegraphics[width=8cm]{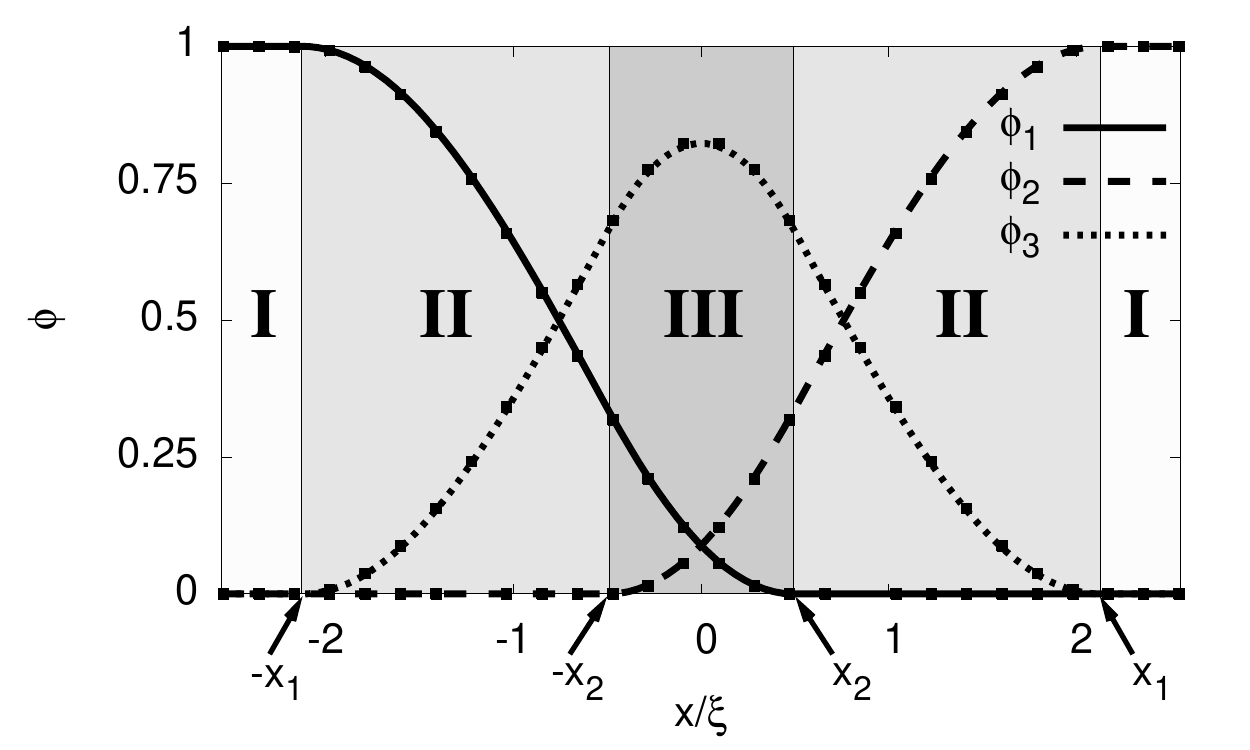}
\caption{Phase field interface profiles in the multi-order parameter case.
The curves result from the analytical calculations, the points from a full numerical minimisation of the free energy functional.
The bold Roman region numbers are used to distinguish between the regions with a different number of nontrivial phase fields and corresponding sets of governing equations.
Parameters are $\sigma_{12}/\sigma_{13}=3$ and $\dimtemp_{13}=0.25$.
}
\label{fig6}
\end{center}
\end{figure}

As we have seen in the analysis in the preceding section, a single order parameter model always leads to an attractive interaction between the solid-melt interfaces.
This is due to the effect that complete disappearance of the melt layer ($\Wbar=0$) leads to a perfectly healed crystal ($\phi\equiv 1$), and therefore interfacial energy contributions disappear completely, which reduces the total energy.
Here, instead, we discuss situations, where the left and right semi-infinite crystal are characterised by different order parameters.
Although we mainly have here in mind to distinguish between two different grain orientations of the same phase, which are separated by its melt, one can also consider more generally segregation phenomena at grain boundaries, and also the extension towards wetting of different phases is possible.
In the following we therefore use $\phi_1$ and $\phi_2$ for the two solid phases and $\phi_3$ for the melt phase, as shown in Fig.~\ref{fig6}.
Since we have seen in the preceding section that different coupling functions and different definitions of the melt layer thickness only lead to quantitative differences but do not change the overall behaviour, we restrict the analysis here to just one case for clarity.
In particular, we choose $g(\phi_1, \phi_2, \phi_3)=1-\phi_3$ for the local volume fraction of the solid phases.
The generalisation to other definitions as above is straightforward.

As shown in Fig.~\ref{fig6} the phase fields separate the entire domain into three different regions.
Here we focus for simplicity on symmetrical grain boundaries, which implies that the solid-melt interfacial energies and thickness are are equal.
In particular, we use $\eta_{\alpha\beta}=\eta$ equal for all interfaces.
For the interfacial energies we assume $\sigma_{sl}=\sigma_{13}=\sigma_{23}$, which differs in general from the grain boundary energy $\sigma_{gb}=\sigma_{12}$.
Then the problem is symmetrical, and we choose the origin of the coordinate system such that it is located at the symmetry point.
In region I one of the solid phase fields $\phi_1$ or $\phi_2$ is equal to one, and the other fields vanish.
In region II the melt phase starts to appear, but one of the phase fields of the solid phases is still zero.
Finally, in region III all phase fields acquire nontrivial values.
In the following the phase field equations will be solved separately in all these regions and matched together.
Due to the symmetry it is sufficient to solve the problem for $x\leq 0$.

First, in region I ($x<-x_1$) trivially $\phi_1^{(I)}=1$ and $\phi_2^{(I)}=\phi_3^{(I)}=0$, where the superscript denotes the region.

Second, in region II ($-x_1<x<-x_2$) the equilibrium is obtained by minimisation of the functional $F_0(\phi_1, 0, 1-\phi_1)$ in Eq.~(\ref{intro::eq1}) with respect to $\phi_1$.
This gives in analogy to Eq.~(\ref{single::eq1})
\begin{equation}
-2\xi^2\phi_1'' + 1-2\phi_1 + \dimtemp_{13} = 0
\end{equation}
with $\dimtemp_{13}=L\eta(T-T_M)/[4\sigma_{13}T_M]$, which is the same measure for the overheating as in section \ref{single::section} (remember that $\sigma_{13}$ is the solid-melt interfacial energy, which was previously denoted by $\sigma$).
The solution of this equation is 
\begin{equation}
\phi_1^{\mathrm{(II)}}(x)= \frac{1+\dimtemp_{13}}{2} + A^{\mathrm{(II)}}\sin\frac{x+x_0^{\mathrm{(II)}}}{\xi}.
\end{equation}
The two parameters $A^{\mathrm{(II)}}$ and $x_0^{\mathrm{(II)}}$ are integration constants for the second order differential equation.

Third, in region III ($-x_2<x<x_2$) the equilibrium conditions follow from variation of the free energy $F_0(\phi_1, \phi_2, 1-\phi_1-\phi_2)$ both with respect to $\phi_1$ and $\phi_2$.
They read explicitly
\begin{eqnarray}
&&1-2\phi_1- \left(2-\frac{\sigma_{12}}{\sigma_{13}} \right) \phi_2 - 2\xi^2\phi_1'' \nonumber \\
&&-\xi^2\left(2-\frac{\sigma_{12}}{\sigma_{13}}\right)\phi_2'' =- \dimtemp_{13} 
\end{eqnarray}
and
\begin{eqnarray}
&&1-2\phi_2 - \left(2-\frac{\sigma_{12}}{\sigma_{13}} \right) \phi_1 - 2\xi^2\phi_2'' \nonumber \\
&&-\xi^2\left(2-\frac{\sigma_{12}}{\sigma_{13}}\right)\phi_1'' = -\dimtemp_{13}.
\end{eqnarray}
The general solution of these coupled equations is
\begin{eqnarray}
\phi_1^{\mathrm{(III)}}(x) &=& \frac{1+\dimtemp_{13}}{4-\frac{\sigma_{12}}{\sigma_{13}}} + A^{\mathrm{(III)}}\sin\frac{x+x_0^{\mathrm{(III)}}}{\xi}, \label{lab1} \\
\phi_2^{\mathrm{(III)}}(x) &=& \frac{1+\dimtemp_{13}}{4-\frac{\sigma_{12}}{\sigma_{13}}} - A^{\mathrm{(III)}}\sin\frac{x-x_0^{\mathrm{(III)}}}{\xi}, \label{lab2} 
\end{eqnarray}
in agreement with the symmetry property $\phi_1(x)=\phi_2(-x)$.

Similar to the considerations in the previous section for the single order parameter model we require continuity of the phase fields at the connecting points $-x_1$ and $-x_2$, but do not a priori demand continuity of the slopes.
Hence we get as boundary conditions
\begin{eqnarray}
\phi_1^\mathrm{(II)}(-x_1) &=& 1, \\
\phi_2^\mathrm{(III)}(-x_2) &=& 0, \\
\phi_1^\mathrm{(II)}(-x_2) &=& \phi_1^\mathrm{(III)}(-x_2).
\end{eqnarray}
The six unknowns of the problem are $x_1$, $x_2$, $x_0^\mathrm{(II)}$, $x_0^\mathrm{(III)}$, $A^\mathrm{(II)}$ and $A^\mathrm{(III)}$.
We have checked numerically that the free energy is minimised if indeed the slopes of the phase fields are also continuous at the connection points, i.e.
\begin{eqnarray}
{\phi_1^\mathrm{(II)}}'(-x_1) &=& 0, \\
{\phi_2^\mathrm{(III)}}'(-x_2) &=& 0, \\
{\phi_1^\mathrm{(II)}}'(-x_2) &=& {\phi_1^\mathrm{(III)}}'(-x_2),
\end{eqnarray}
which provides the remaining equations to determine the unknowns.
The solution for these coefficients is
\begin{eqnarray}
x_0^\mathrm{(II)} &=&  x_1 + \frac{\pi \xi}{2}, \quad
A^\mathrm{(II)} = \frac{1-\dimtemp_{13}}{2}, \\
x_0^\mathrm{(III)} &=& -x_2 - \frac{\pi\xi}{2}, \quad
A^\mathrm{(III)} = \frac{1+\Delta_{13}}{4-\sigma_{12}/\sigma_{13}},
\end{eqnarray}
and $x_1$ and $x_2$ are determined by the relations
\begin{equation}
A^\mathrm{(II)}\sin\frac{x_1-x_2}{\xi} = A^\mathrm{(III)} \sin\frac{2x_2}{\xi}
\end{equation}
and
\begin{eqnarray}
&& \frac{1+\dimtemp_{13}}{2} + A^\mathrm{(II)}\cos\frac{x_1-x_2}{\xi} \nonumber \\
&=& \frac{1+\Delta_{13}}{4-\sigma_{12}/\sigma_{13}}- A^\mathrm{(III)} \cos\frac{2x_2}{\xi}.
\end{eqnarray}
We mention in passing that physically reasonable kink-free solutions exist only for $\sigma_{12}/\sigma_{13}<4$, see Eqs.~(\ref{lab1}) and (\ref{lab2}).
Higher ratios are therefore not considered in the following.

\begin{figure}
\begin{center}
\includegraphics[width=8cm]{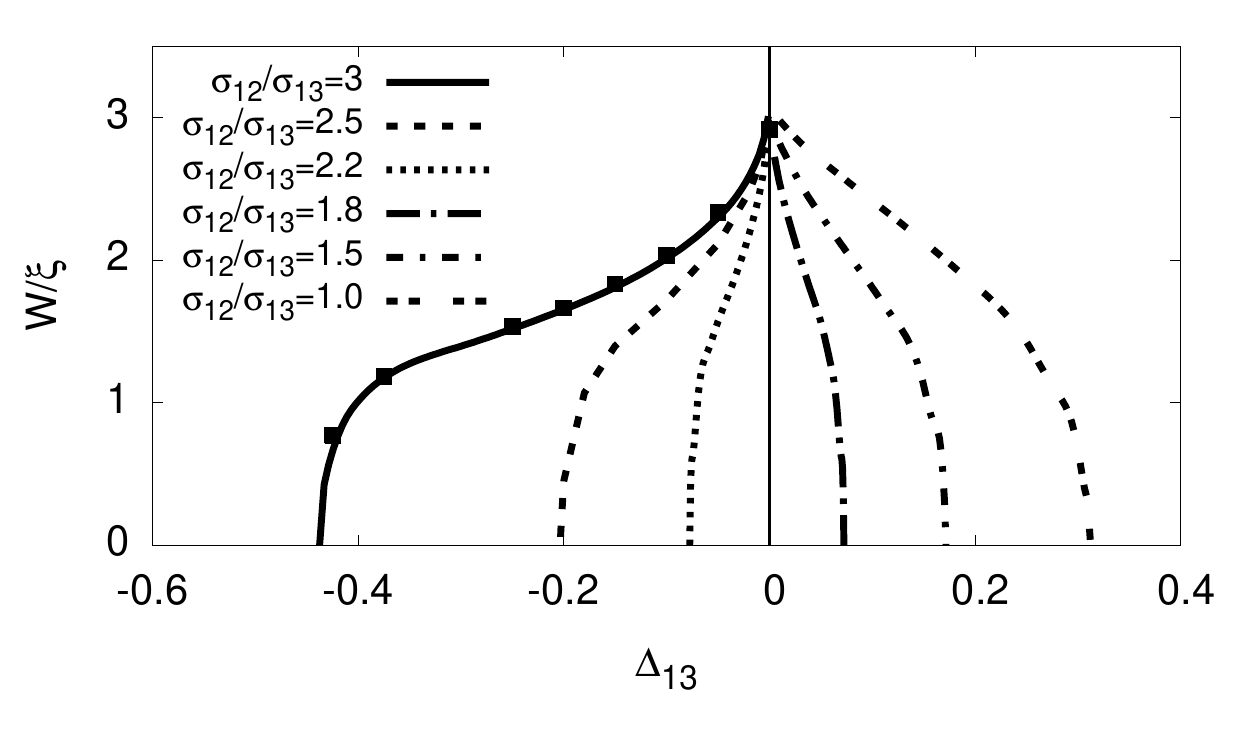}
\caption{Melt layer thickness as function of temperature for different ratios of the interfacial energies. 
The curves follow from the analytical solution for different values of $\sigma_{12}/\sigma_{13}$.
For the case $\sigma_{12}/\sigma_{13}=3$ also results from an independent full numerical solution are shown as squares.
They agree perfectly with the analytical solution.
The transition between attractive and repulsive situations occurs for $\sigma_{12}=2\sigma_{13}$.
}
\label{fig8}
\end{center}
\end{figure}

From the preceding results the melt layer thickness can be computed, and this is shown in Fig.~\ref{fig8}.
As expected, we find now a transition from attractive to repulsive interaction at $\sigma_{12}=2\sigma_{13}$.
A characteristic feature is the inflection point in all curves, which is due to the fact that depending on the temperature, the point where the melt phase fraction is below $\phi_3=1/2$ can be located either in region II or III in Fig.~\ref{fig6}.

\begin{figure}
\begin{center}
\includegraphics[width=8cm]{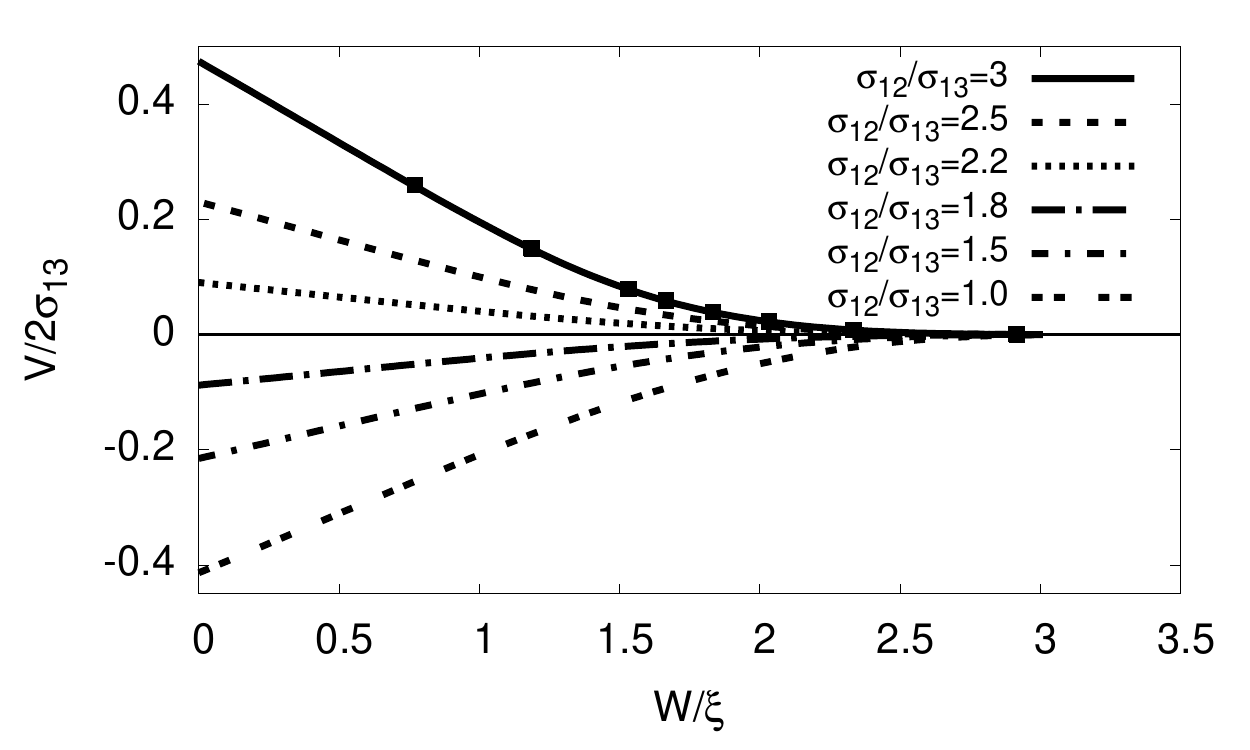}
\caption{Disjoining potential as function of the melt layer thickness for the multi-order parameter phase field model.
For $\sigma_{12}/\sigma_{13}=3$ the analytical data is confirmed by fully numerical results (black squares).
}
\label{fig9}
\end{center}
\end{figure}

From the corresponding total free energy per unit area of the system $\cal F$  we can then determine the disjoining potential according to
\begin{equation} \label{eq2}
\frac{V(W)}{2\sigma_{13}} = \frac{{\cal F}}{2\sigma_{13}} + \frac{2}{\pi} \dimtemp_{13} \frac{W}{\xi} - 1.
\end{equation}
The result is shown in Fig.~\ref{fig9}.
Here it is directly visible that the transition between purely attractive and repulsive interactions takes place at the classical threshold $\sigma_{12}/\sigma_{13}=2$.
The interaction becomes strictly zero for $W/\xi>\pi$.

\begin{figure}
\begin{center}
\includegraphics[width=8cm]{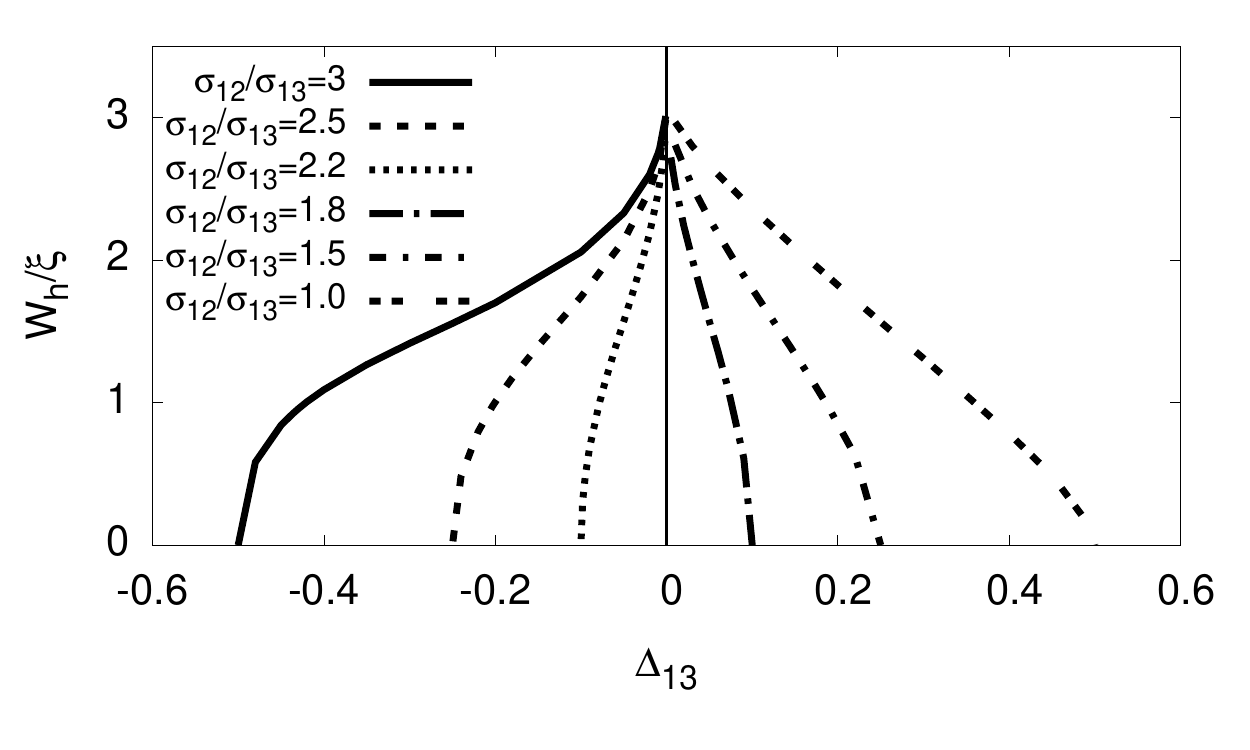}
\caption{Melt layer thickness as function of temperature, using the integral measure $\Wbar$ according to Eq.~(\ref{eq1}).
For lower temperatures than those, where the curves hit the horizontal axis, a dry grain boundary with $\phi_3\equiv 0$ (thus $\Wbar=0$) is the equilibrium solution.
}
\label{fig10}
\end{center}
\end{figure}

In agreement with the definition (\ref{single::eq2}) in the single order parameter case the integral definition of the melt layer thickness reads here
\begin{equation} \label{eq1}
\Wbar=\int_{-\infty}^\infty \phi_3\,dx.
\end{equation}
The equilibrium melt layer thickness as function of temperature using this measure is shown in Fig.~\ref{fig10}.
These curves look very similar to the ones in Fig.~\ref{fig8}, indicating that here the choice of the measure for the melt layer thickness is not important.

\begin{figure}
\begin{center}
\includegraphics[width=8cm]{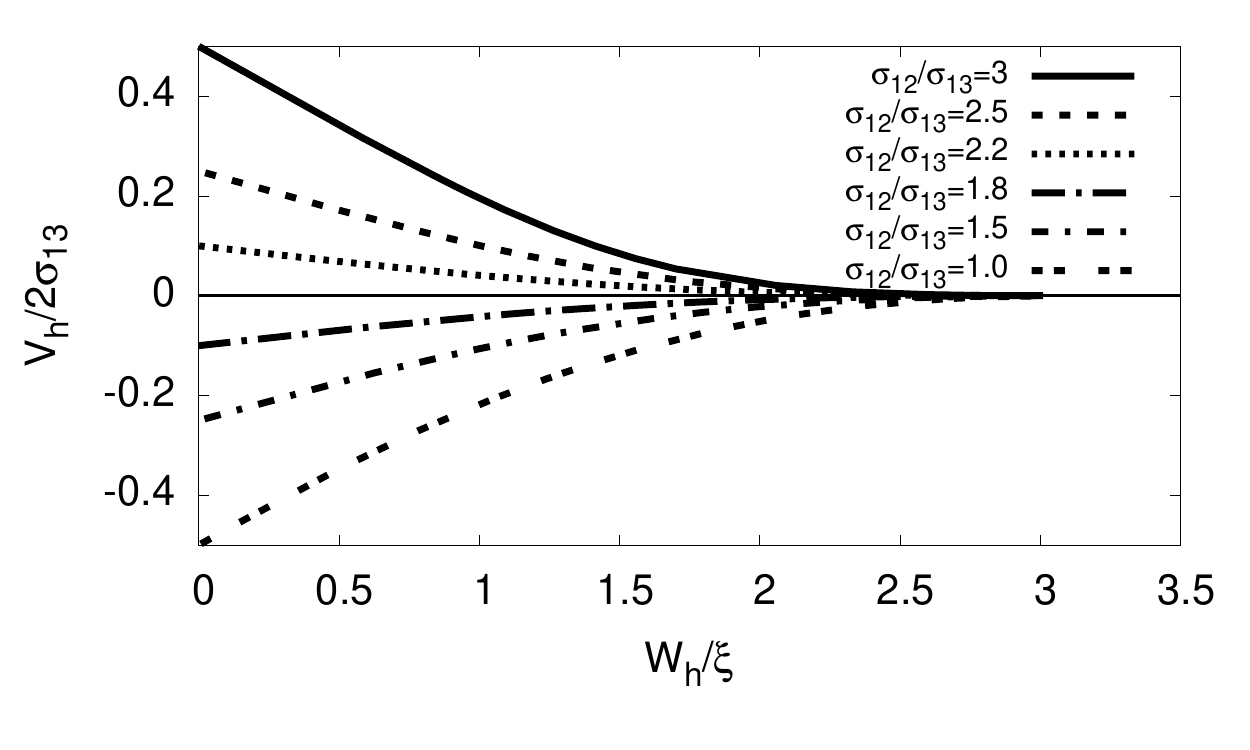}
\caption{Disjoining potential $\Vbar$ as function of the melt layer thickness $\Wbar$ for the multi-order parameter phase field model.
}
\label{fig11}
\end{center}
\end{figure}

Similar to Eq.~(\ref{eq2}) we can then determine also the disjoining potential $\Vbar$, which is shown in Fig.~\ref{fig11}.
In the limit $\sigma_{12}/\sigma_{13}\to 0$, i.e.~vanishing grain boundary energy, the behaviour reduces to the single order parameter case, see Fig.~\ref{fig4}.

\begin{figure}
\begin{center}
\includegraphics[width=8cm]{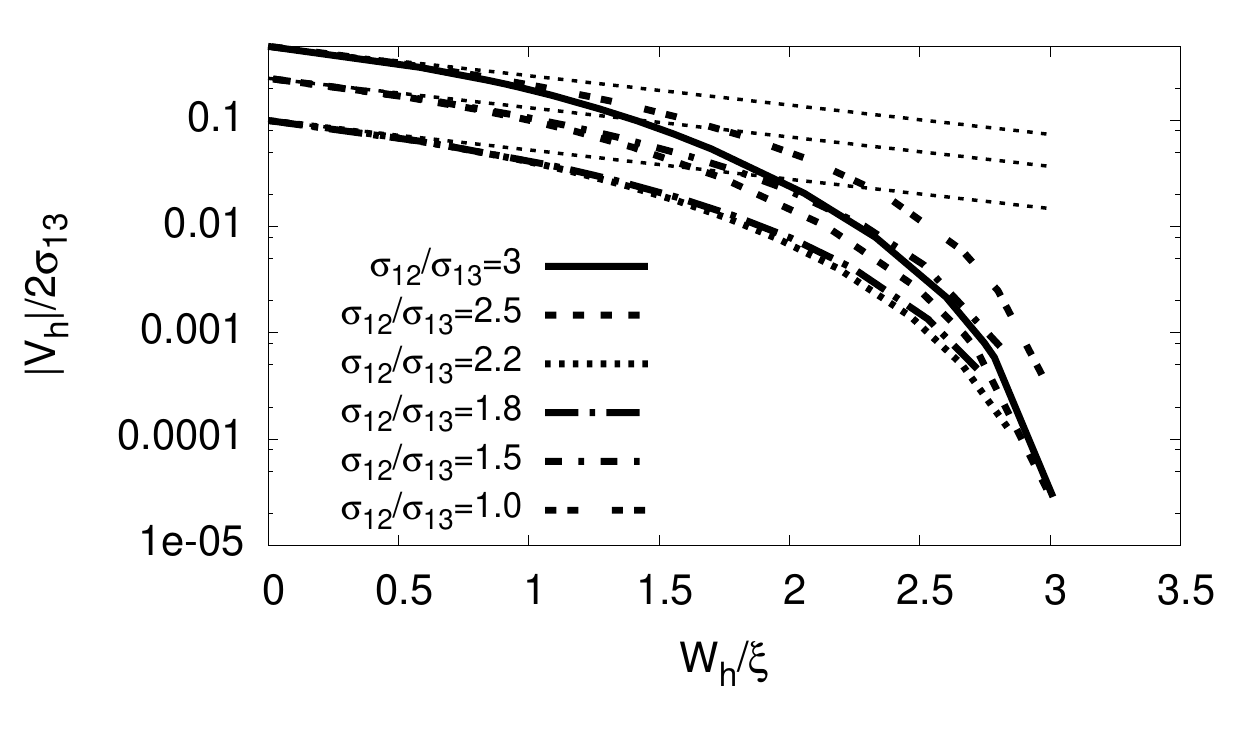}
\caption{Absolute value of the disjoining potential $\Vbar$ as function of the melt layer thickness $\Wbar$ for the multi-order parameter phase field model.
The dotted straight lines correspond to the empirical expression (\ref{ana}) and agree with the multi-order parameter phase field result for small values of $\Wbar$.
}
\label{fig13}
\end{center}
\end{figure}

To better understand the transition between the attractive and repulsive situations, we can look at the {\em bridging temperature} $\Delta_b$, at which the melt layer thickness $\Wbar$ becomes zero.
In the free energy landscape (\ref{intro::eq1}) it is sufficient to inspect the local terms, which appear in the free energy density for our choice of the coupling function as
\begin{eqnarray}
f_l(\phi_1, \phi_2) &=& \frac{4\sigma_{12}}{\eta} \phi_1\phi_2 + \frac{4\sigma_{13}}{\eta} \phi_1(1-\phi_1-\phi_2) \nonumber \\
&+& \frac{4\sigma_{23}}{\eta} \phi_2(1-\phi_1-\phi_2) \nonumber \\
&-& L \frac{T-T_M}{T_M}(1-\phi_1-\phi_2),
\end{eqnarray}
where we immediately set $\phi_3=1-\phi_1-\phi_2$.
A dual interface between the solid phases can be free of a wetting melt phase if any appearance of $\phi_3>0$ increases the energy.
In the $(\phi_1, \phi_2)$ plane the dual interface between phase 1 and phase 2 corresponds to the diagonal $\phi_1+\phi_2=1$, and the orthogonal direction $\vec{n}=(-1,-1)$ to the appearance of the third phase 3.
The criterion for wetting ($\Wbar>0$) is then $\nabla f_l\cdot\vec{n}<0$.
The threshold, where this inner product vanishes, gives therefore the bridging temperature $\Delta_b$.
Evaluation gives for $\sigma_{13}=\sigma_{23}$
\begin{equation}
\Delta_b = \frac{1}{2} \left( 2-\frac{\sigma_{12}}{\sigma_{13}} \right),
\end{equation}
which agrees with the results in Fig.~\ref{fig10}.
First, we see that the bridging temperature changes sign exactly at the classical threshold for grain boundary wetting, $\sigma_{12}=2\sigma_{13}$.
Second, this result is in agreement with the classical expectation \cite{Rappaz:2003aa}
\begin{equation}
\frac{T_b-T_M}{T_M} = -\frac{\sigma_{gb}-2\sigma_{sl}}{L\delta},
\end{equation}
with the lengthscale $\delta = \eta/2$.
This prediction is based on the shape of the disjoining potential
\begin{equation} \label{ana}
V(W) = \bar{\sigma} \exp(-W/\delta)
\end{equation}
with $ \bar{\sigma}=\sigma_{12}-2\sigma_{13}$.
A comparison with the actually determined disjoining potentials $\Vbar(\Wbar)$ is shown in Fig.~\ref{fig13} in a semi-logarithmic representation.
The results for the multi-order parameter phase field model do not have the exponential structure
since the interaction between interfaces must end at a finite separation with a multi-obstacle potential.
However, for low values of $\Wbar$ it agrees well with the empirical model (\ref{ana}) using the above value for the interaction range $\delta=\eta/2$.
%

\section{Summary and conclusions}

We have analysed the ability of the widely used phase field model \cite{Steinbach:1998aa} to describe grain boundary premelting.

In the single order parameter case we find always an attractive interaction between overlapping solid-melt interfaces.
This is due to the fact that merging of the solid-phases removes the interfacial energy between the two grains with identical orientation, which are represented by the same order parameter.
The interaction vanishes as soon as the interfaces with finite thickness do not overlap anymore.
The precise predictions for the melt layer thickness and the disjoining potential depend on the choice of the thermal coupling function and the definition of the melt film width.

In the multi-order parameter case we find both attractive and repulsive interactions, and the transition between them appears at the classically expected threshold $\sigma_{sl}/\sigma_{gb}=2$.
In contrast to phase field models with a multi-well potential the interaction is strictly repulsive above this transition and does not exhibit an attractive tail, which was considered to be not physical \cite{PhysRevE.81.051601}.
Instead, the interaction vanishes completely as soon as the distance $W$ between the solid-melt interfaces exceeds $\eta=\pi\xi$, which is the case when the interfaces no longer overlap.
Consequently, due to this cutoff the disjoining potential differs also from a phenomenological exponential decay at large distances, which has also been found in amplitude equations and phase field crystal descriptions.
The disjoining potential is either attractive or repulsive, in agreement with the empirical analytical model, while the disjoining potentials predicted by Molecular Dynamics, phase field crystal and amplitude equations simulations exhibit a shallow minimum at a finite width in the attractive case.
In this sense the multi-order parameter based phase field model does not reflect that a grain boundary retains a finite amount of disorder at the melting point.

From a practical point of view, the multi-order parameter phase field model therefore captures grain boundary premelting at least qualitatively.
We point out that a rather fine discretisation of the interface is needed in the simulations in order to obtain accurate interactions.
This is to a significant extent attributed to the piecewise definition of the free energy functional and the derived equilibrium conditions and equations of motion.

\section*{Acknowledgements}

This work has been supported by the Deutsche Forschungsgemeinschaft via the Collaborative Research Center 761 {\em Steel ab initio}.
The work of A.K. was supported by the US DOE grant DEFG02-07ER46400.
I.S. would like to acknowledge the support of the Deutsche Forschungsgemeinschaft via the Collaborative Research Center TR103 {\em Superalloy Single Crystals}.

\section*{References}

\bibliographystyle{elsarticle-num} 
\bibliography{references}





\end{document}